\documentclass[copyright]{eptcs}
\providecommand{\event}{I. McQuillan, G. Pighizzini (Eds.): 12th
International Workshop\\
on Descriptional Complexity of Formal Systems (DCFS 2010)} 

\usepackage{amssymb,amsmath}
\usepackage[dvips]{graphics}
\usepackage{epsfig}

\usepackage{latexsym}

\def\bbbn{{\rm I\!N}} 

\def \prend{\vrule depth-1pt height7pt width6pt}
\def \proof{\bigbreak\noindent{\bf Proof.\ \ }}
\def \endpf{{\ \ \prend \medbreak}}

\begin{document}

\renewcommand{\labelenumi}{(\roman{enumi})}

\if01
\newtheorem{theorem}{T\/heorem}[section]
\newtheorem{definition}{Definition}[section]
\newtheorem{lemma}{Lemma}[section]
\newtheorem{applem}{Lemma}[section]
\newtheorem{example}{Example}[section]
\newtheorem{comment}{Comment}[section]
\newtheorem{remark}{Remark}[section]
\newcommand{\propersubset}{\subset}
\newtheorem{open}{Open problem}[section]
\fi
\newtheorem{theorem}{Theorem}
\newtheorem{corollary}{Corollary}
\newtheorem{lemma}{Lemma}
\newtheorem{proposition}{Proposition}
\newtheorem{open}{Open problem}
\newtheorem{conjecture}{Conjecture}

\title{Transition Complexity of Incomplete DFAs
\thanks{Research supported in part by the Natural Sciences and
Engineering Research Council of Canada. All correspondence should be
directed to Sheng Yu at syu@csd.uwo.ca.}}

\author{Yuan Gao
\institute{Department of Computer Science, \\
The University of Western Ontario,\\
London, Ontario, Canada N6A 5B7} \email{ygao72@csd.uwo.ca} \and Kai
Salomaa
\institute{School of Computing, \\
Queen's University, \\
Kingston, Ontario, Canada K7L 3N6} \email{ksalomaa@cs.queensu.ca}
\and Sheng Yu
\institute{Department of Computer Science, \\
The University of Western Ontario,\\
London, Ontario, Canada N6A 5B7} \email{syu@csd.uwo.ca}}

\def\titlerunning{Transition Complexity of Incomplete DFAs}
\def\authorrunning{Y. Gao, K. Salomaa, S. Yu}


\maketitle

\begin{abstract}
In this paper, we consider the transition complexity of regular languages based on
the incomplete deterministic finite automata.
A number of results on Boolean operations
have been obtained.
It is shown that the transition complexity results for union and complementation
are very different from the state complexity results for the same operations.
However, for intersection, the transition complexity result is similar to that
of state complexity.
\end{abstract}

\section{Introduction}

Many results have been obtained in recent years on the state
complexity of individual and combined operations of regular
languages and a number of sub-families of regular languages
\cite{DoOk09,Jriaskova05,PiSh2002,SaSaYu07,Yu01,Yu2}. The study of
state complexity has been mostly based on the model of complete
deterministic finite automata (DFAs). When the alphabet is fixed,
the number of states of a complete DFA determines the number of
transitions of the DFA. Note that a description of a DFA consists of
a list of transitions, which determine the size of the DFA.
Incomplete DFAs are implied in many publications~\cite{Ca,Wo87}. In
quite a number of applications of finite automata, incomplete rather
than complete DFAs are more suitable for those applications
\cite{Karttunen,Watt}. For example, in natural language and speech
processing, the input alphabet of a DFA commonly includes at least
all the ASCII symbols or the UNICODE symbols. However, the number of
useful transitions from each state is usually much smaller than the
size of the whole alphabet, which may include only a few symbols
\cite{Karttunen}. Although the state complexity of such an
incomplete DFA can still give a rough estimate of the size of the
DFA, the number of transitions would give a more precise measurement
of its size.

In this paper, we consider the descriptional complexity measure
that counts the number of transitions in an incomplete DFA.
It is clear that for two DFAs with an equal number of states, the
size of the description may be much smaller for a DFA where many
transitions are undefined. Especially, for applications that use
very large or possibly non-constant alphabets, or DFAs with most
transitions undefined, it can be argued that transition complexity
is a more accurate descriptional complexity measure than state
complexity. Before this paper, transition complexity was
investigated only on nondeterministic finite
automata~\cite{DoSa07,DoSa08,GrHo07,HoKu09} and Watson-Crick finite
automata~\cite{PaPa99}.

We consider operational transition complexity of Boolean operations.
The transition complexity results for union and
complementation turn out to be essentially different from
the known state complexity results~\cite{Yu2,Yu} that deal with complete DFAs.
Perhaps, as expected, the results for intersection are more similar
with the state complexity results.
For union we have upper and lower bounds that differ, roughly, by a multiplicative constant of 2.
We conjecture that worst-case examples for transition
complexity of union need to be based on two DFAs that
for some symbol of the alphabet have all transitions defined.
For DFAs of this type we have a tight transition complexity bound for union.

We can note that for union the state complexity results
are also different for complete and incomplete DFAs, respectively.
When dealing with incomplete DFAs
the state complexity of the union
 of an $n_1$ state and an $n_2$ state language is in the worst case
$n_1 \cdot n_2 + n_1 + n_2$.

\section{Preliminaries}

In the following, $\Sigma$ denotes a finite alphabet, $\Sigma^*$ is
the set of strings over $\Sigma$ and $\varepsilon$
is the empty string. A language is any subset of $\Sigma^*$.
When $\Sigma$ is known, the complement of a language
$L \subseteq \Sigma^*$ is denoted as $L^c = \Sigma^* - L$.

A deterministic finite automaton (DFA) is a tuple $A = (\Sigma, Q,
q_0, F, \delta)$ where $\Sigma$ is the input alphabet, $Q$ is the
finite set of states, $F \subseteq Q$ is the set of accepting states
and the transition function $\delta$ is a partial function $Q \times
\Sigma \rightarrow Q$. The transition function is extended in the
usual way to a (partial) function $\hat{\delta} : Q \times \Sigma^*
\rightarrow Q$ and also $\hat{\delta}$ is denoted simply by
$\delta$. The language recognized by $A$ is $L(A) = \{ w \in
\Sigma^* \mid \delta(q_0, w) \in F \}$.

Unless otherwise mentioned, by a DFA we mean always  an incomplete
DFA, that is, some transitions may be undefined. For more knowledge
in incomplete automata, the reader may refer to~\cite{Ca}. The state
complexity of a regular language $L$, ${\rm sc}(L)$, is the number
of states of the minimal incomplete DFA recognizing $L$.

The Myhill-Nerode right congruence of a regular language $L$ is
denoted $\equiv_L$~\cite{Yu}. The number of equivalence classes of
$\equiv_L$ is equal to ${\rm sc}(L)$ if the minimal DFA for $L$ has
no undefined transitions, ${\rm sc}(L)+1$ otherwise.



If $A = (\Sigma, Q, q_0, F, \delta)$ is as above, the number of
transitions of $A$ is the cardinality of the domain of
$\delta$,
$|{\rm dom}(\delta)|$. In the following the
number of transitions of $A$ is denoted
$\#_{\rm tr}(A)$. We note that if $A$
is connected (that is, all states are reachable
from the start state), then
\begin{equation}
\label{esat1}
|Q| - 1 \leq \#_{\rm tr}(A)  \leq |\Sigma| \cdot |Q|.
\end{equation}
For $b \in \Sigma$, the number of transitions
labeled by $b$ in $A$
 is denoted $\#_{\rm tr}(A, b)$.

The {\em transition complexity\/} of a regular language $L$, ${\rm tc}(L)$,
is the minimal number of transitions of any DFA recognizing $L$.
In constructions establishing bounds for the number of transitions
it is sometimes useful to restrict consideration to transitions
corresponding to a particular alphabet symbol and we introduce
the following notation.
For $b \in \Sigma$, the {\em $b$-transition complexity of $L$,}
${\rm tc}_b(L)$ is the minimal number of $b$-transitions
of any DFA recognizing $L$. The following lemma establishes that
for any $b \in \Sigma$, the state minimal DFA for $L$ has
 the minimal number of $b$-transitions of any DFA recognizing $L$.

\begin{lemma}
\label{mini}
Suppose that $A
= (\Sigma, Q, q_0, F, \delta)$ is the state minimal DFA for
a language $L$.
For any $b \in \Sigma$,
$$
{\rm tc}_b(L) = \#_{\rm tr}(A, b).
$$
\end{lemma}



Since the result is expected, we omit the proof. Lemma~\ref{mini}
means, in particular, that for any given $b \in \Sigma$ we cannot
reduce the number of $b$-transitions by introducing additional
states or transitions for other input symbols. From Lemma~\ref{mini}
it follows that
$${\rm tc}(L) = \sum_{b \in \Sigma}{\rm tc}_b(L).
$$
As a corollary of Lemma~\ref{mini} we have also:

\begin{corollary}
\label{cor1}
Let $A$ be the minimal DFA for a language $L$. For any
$b \in \Sigma$, the number of undefined $b$-transitions
in $A$ is ${\rm sc}(L) - {\rm tc}_b(L)$.
\end{corollary}

To conclude this section, we give a formal asymptotic definition of  the
transition complexity of an operation on regular languages.
Let $\odot$ be an $m$-ary operation, $m \geq 1$, on languages and let
$f : \bbbn^m \rightarrow \bbbn$. We say that the transition complexity
of $\odot$ is $f$ if
\begin{description}
\item[{\rm (a)}] for all regular languages $L_1$, \ldots, $L_m$,
\begin{equation}
\label{sce} {\rm tc}(\odot(L_1, \ldots, L_m)) \leq f({\rm tc}(L_1),
\ldots, {\rm tc}(L_m)),
\end{equation}
\item[{\rm (b)}] for any $(n_1, \ldots, n_m) \in \bbbn^m$ there
exist $n_i' \geq n_i$, $i = 1, \ldots, m$,
and regular languages $L_i$ with ${\rm tc}(L_i) = n_i'$,
$i = 1, \ldots, m$, such that the equality holds in~(\ref{sce}).
\end{description}
The above definition requires that there exist
worst case examples
with arbitrarily large transition complexity
matching the upper bound, however, we do not require that
matching lower bound examples exist where the argument
languages have transition
complexity exactly $n_i$ for all positive integers $n_i$,
$i = 1, \ldots, m$.

\section{Transition complexity of union}

We first give bounds for the number
of transitions corresponding to a particular input symbol $b \in \Sigma$.
These bounds will be used in the next subsection to develop
 upper bounds for the total number of transitions
needed to recognize the union of two languages.

\subsection{Number of transitions corresponding to a fixed
 symbol}

The upper bound for
the $b$-transition complexity of union of languages
$L_1$ and $L_2$ depends also
on the number of states of the minimal
DFAs for $L_i$, $i = 1, 2$,  for which a $b$-transition is not
defined. From Corollary~\ref{cor1} we recall that this quantity
equals to ${\rm sc}(L_i) - {\rm tc}_b(L_i)$.

\begin{lemma}
\label{lem1}
Suppose that $\Sigma$ has at least two symbols and
 $L_1$, $L_2$ are regular languages over $\Sigma$. For
any $b \in \Sigma$,
\begin{eqnarray}
\label{ubound}
{\rm tc}_b(L_1 \cup L_2) & \leq  &
 {\rm tc}_b(L_1)
 \cdot {\rm tc}_b(L_2) + \nonumber \\
& &  {\rm tc}_b(L_1)(1 + {\rm sc}(L_2) - {\rm tc}_b(L_2))
+  {\rm tc}_b(L_2)(1 + {\rm sc}(L_1) - {\rm tc}_b(L_1)).
\end{eqnarray}
If $n_1, n_2 \geq 2$  are relatively prime,
for any $1 \leq k_i < n_i$, $i = 1, 2$, there exist regular languages
$L_i$ with ${\rm sc}(L_i) = n_i$,
and ${\rm tc}_b(L_i) = k_i$, $i = 1, 2$, such that the inequality
(\ref{ubound}) is an equality.
\end{lemma}

\proof
Consider regular languages $L_i$ and let $A_i = (\Sigma,
Q_i, q_{0,i}, F_i, \delta_i)$ be a DFA
 recognizing
$L_i$, $i = 1, 2$. From $A_1$ and
$A_2$ we obtain a DFA for
$L_1 \cup L_2$ using the well-known
cross-product construction~\cite{Yu} modified
to the case of incomplete automata.
We define
$$
Q_i' = \left\{
\begin{array}{l}
Q_i \cup \{ d \} \mbox{ if $A_i$ has some undefined transitions, }\\
Q_i, \mbox{ otherwise,}
\end{array} \right.
\;\; i = 1, 2.
$$
Now let
\begin{equation}
\label{uconst}
B = (\Sigma, Q_1' \times Q_2', (q_{0,1}, q_{0,2}), (F_1 \times Q_2')
\cup (Q_1' \times F_2),
\gamma),
\end{equation}
where for $b \in \Sigma$, $q_i' \in Q_i'$, $i = 1, 2$,
\begin{equation}
\label{latta2}
\gamma((q_1', q_2'), b) = \left\{
\begin{array}{l}
(\delta_1(q_1', b), \delta_2(q_2', b)) \mbox{ if }
\delta_1(q_1', b) \mbox{ and } \delta(q_2', b)
\mbox{ are both defined,}\\
(\delta_1(q_1', b), d) \mbox{ if } \delta_1(q_1', b) \mbox{ is defined and }
\delta_2(q_2', b) \mbox{ is undefined,}\\
(d, \delta_2(q_2', b)) \mbox{ if } \delta_1(q_1', b) \mbox{ is undefined and }
\delta_2(q_2', b) \mbox{ is defined,}\\
\mbox{undefined, otherwise.}
\end{array} \right.
\end{equation}
Note that above $\delta_i(d, b)$, $i = 1, 2$,  is
always undefined for any $b \in \Sigma$.

We note that for $b \in \Sigma$,
\begin{eqnarray}
\label{latta3}
\#_{\rm tr}(B, b) & = & \#_{\rm tr}(A_1, b) \cdot \#_{\rm tr}(A_2, b)
+ \#_{\rm tr}(A_1, b) +  \#_{\rm tr}(A_2, b) + \nonumber \\
& & \#_{\rm tr}(A_1, b) \cdot (|Q_2| - \#_{\rm tr}(A_2, b))
+ \#_{\rm tr}(A_2, b) \cdot (|Q_1| - \#_{\rm tr}(A_1, b)).
\end{eqnarray}
Here
\begin{itemize}
\item $\#_{\rm tr}(A_1, b) \cdot \#_{\rm tr}(A_2, b)$
is the number of transitions in~(\ref{latta2}) where
both $\delta_i(q_i', b)$, $i = 1, 2$, are defined,
\item $\#_{\rm tr}(A_i, b) \cdot (|Q_j| - \#_{\rm tr}(A_j, b))$,
 $\{ i, j \} = \{ 1, 2 \}$, is the number of transitions
in~(\ref{latta2}) where $\delta_i(q_i', b)$ is defined,
$\delta_j(q_j', b)$ is undefined and $q_j' \in Q_j$, and,
\item $\#_{\rm tr}(A_i, b)$ is the number of transitions
in~(\ref{latta2}) where $\delta_i(q_i', b)$ is defined and
$q_j' = d$, $\{ i, j \} = \{ 1, 2 \}$.
\end{itemize}
By choosing $A_i$ as the minimal DFA for $L_i$, $i = 1, 2$,
and using Lemma~\ref{mini} and Corollary~\ref{cor1},
the right side of equation~(\ref{latta3}) gives
the right side of inequality~(\ref{ubound}). Since
$B$ recognizes $L_1 \cup L_2$,
${\rm tc}_b(L_1 \cup L_2) \leq \#_{\rm tr}(B, b)$.

We give a construction for the lower bound.
Fix $b \in \Sigma$.
Let $n_1, n_2 \geq 1$ be relatively prime,
$1 \leq k_i < n_i$, $i = 1, 2$,  and
 let $c \in \Sigma$ be a symbol distinct from $b$.
Define
$$C_i = (\Sigma, \{ q_{0,i}, q_{1,i}, \ldots, q_{n_i - 1, i} \},
q_{0,i}, \{ q_{0,i} \}, \delta_i),
$$
where the transitions defined by $\delta_i$ are as follows:
\begin{itemize}
\item $\delta_i(q_{j,i}, c) = q_{j+1, i}$, $j = 0, \ldots, n_i -2$,
\item $\delta_i(q_{n_i - 1, i}, c) = q_{0, i}$,
\item $\delta_i(q_{j, i}, b) = q_{j, i}$, $j = 0, \ldots, k_i - 1$.
\end{itemize}
\begin{figure}[ht]
  \centering
  \includegraphics[scale=0.60]{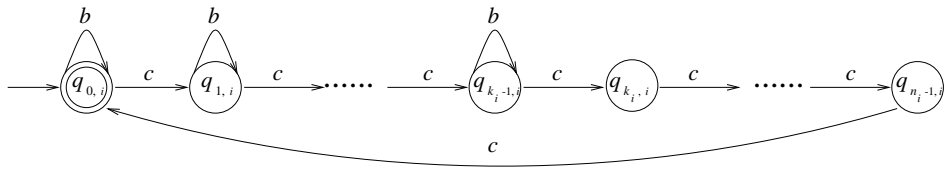}
  \caption{The transition diagram of the witness DFA $C_i$ of Lemma~\ref{lem1}}
\label{union-Ci}
\end{figure}
The transition diagram of $C_i$ is shown in Figure~\ref{union-Ci}.
We note that $\#_{\rm tr}(C_i, b) = k_i$ and
$$
L(C_i) = ((b^*c)^{k_i}c^{n_i-k_i})^*b^*, \;\; i = 1, 2.
$$
Clearly $C_i$ is minimal and hence ${\rm sc}(L(C_i)) = n_i$, $i = 1, 2$.
In the following we denote $L_i = L(C_i)$, $i = 1, 2$, for short.
Choose $m_1, m_2 \in \bbbn$ such that
$$
m_1 \equiv 0 \; ({\rm mod} \; n_1), \;
m_1 \equiv -1 \; ({\rm mod} \; n_2), \;\;
m_2 \equiv 0 \; ({\rm mod} \; n_2), \;
m_2 \equiv -1 \; ({\rm mod} \; n_1).
$$
Since $n_1$ and $n_2$ are relatively prime, the numbers $m_i$, $i = 1, 2$,
exist. The intuitive idea is that we want that the string $c^{m_1}$ takes
the automaton $C_1$ to a state where the $b$-transition is defined and
the same string $c^{m_1}$ takes the automaton $C_2$ to a state where
the $b$-transition is not defined. Recall that $k_2 < n_2$ and
$\delta_2(q_{n_2 - 1, 2}, b)$ is undefined.
Also, a similar property holds
for $c^{m_2}$ with $C_1$ and $C_2$ interchanged.

We define $S = S_1 \cup S_2 \cup S_3$, where $S_1 = \{ c^i \mid 0
\leq i < n_1 \cdot n_2 \}$, $S_2 = \{ c^{m_1} b c^i \mid 0 \leq i <
n_1 \}$ and $S_3 = \{ c^{m_2} b c^i \mid 0 \leq i <  n_2 \}$.
We verify that all strings of $S$ are pairwise in different
equivalence classes of the right congruence $\equiv_{L_1 \cup L_2}$.
First consider $c^i, c^j \in S_1$, $0 \leq i < j < n_1 \cdot n_2$.
Since $n_1$ and $n_2$ are relatively prime, there exists $k \in \{
1, 2 \}$ such that $n_k$ does not divide $j - i$. Denote by $k'$ the
element of $\{ 1, 2 \}$ distinct from $k$. Select $z \in \bbbn$ such
that $i + z \equiv 0 \; ({\rm mod} \; n_k)$ and $j + z \equiv 1 \;
({\rm mod} \; n_{k'})$. Now $c^i c^z \in L_k$ and $c^j c^z \not\in
L_1 \cup L_2$. ($j + z \not\equiv 0 \; {\rm mod} \; n_k$ because
$n_k$ does not divide $j-i$.)

Consider $c^{m_1} b c^i, c^{m_1} b c^j \in S_2$,
$0 \leq i < j <  n_1$.
We note that strings of $S_2$ are not prefixes of any string in $L_2$ and
hence for $z \in \bbbn$ such that $m_1 + i + z \equiv 0 \; ({\rm mod}
\; n_1)$ we have $c^{m_1} b c^i c^z \in L_1$ and
$c^{m_1} b c^j c^z \not\in L_1 \cup L_2$. Similarly we see that
any two elements of $S_3$ are not in the same $\equiv_{L_1 \cup L_2}$-class.

Next consider $c^i \in S_1$ and $c^{m_1} b c^j \in S_2$,
$0 \leq i < n_1 \cdot n_2$, $0 \leq j < n_1$. Choose
$z \in \bbbn$ such that $i + z \equiv 0 \; ({\rm mod} \; n_2)$
and $m_1 + j + z \equiv 1 \; ({\rm mod} \; n_1)$. Now
$c^i c^z \in L_2$ and, since no string
of $S_2$ is a prefix of a string of $L_2$, it follows that
$c^{m_1} b c^j c^z \not\in L_1 \cup L_2$. Completely similarly it follows
that a string of $S_1$ is not equivalent with any string of $S_3$.

As the last case consider $c^{m_1} b c^i \in S_2$ and
$c^{m_2} b c^j \in S_3$, $0 \leq i < n_1$, $0 \leq j < n_2$.
Choose $z \in \bbbn$ such that
$m_1 + i + z \equiv 0 \; ({\rm mod} \; n_1)$ and
$m_2 + j + z \equiv 1 \; ({\rm mod} \; n_2)$. Now
$c^{m_1} b c^i c^z \in L_1$ and, since no string of $S_3$
is a prefix of a string in $L_1$, $c^{m_2} b c^j c^z \not\in L_1 \cup L_2$.

Now we are ready to give a lower bound for the $b$-transition
complexity of $L_1 \cup L_2$.
Let $D$ be the minimal DFA for $L_1 \cup L_2$.
 By Lemma~\ref{mini}
we know that $\#_{\rm tr}(D, b) = {\rm tc}_b(L_1 \cup L_2)$.

For $w \in S$, let $q_w$ be the state of $D$ corresponding to $w$.
We have verified that $q_{w_1} \neq q_{w_2}$ when $w_1 \neq w_2$.
For $c^i \in S_1$,  $0 \leq i < n_1 \cdot n_2$, the string $c^i b$
is a prefix of some string in $L_1 \cup L_2$ if and only if
\begin{equation}
\label{matta1}
i \equiv j \; ({\rm mod} \; n_x) \mbox{ for some
$0 \leq j < k_x$ and some $x \in \{ 1, 2 \}$.}
\end{equation}
The number of integers $0 \leq i < n_1 \cdot n_2$ that satisfy~(\ref{matta1})
with value $x \in \{ 1, 2 \}$ is equal to $k_x \cdot n_y$,
where $\{ x, y \} = \{ 1, 2 \}$, and the number
of integers $0 \leq i < n_1 \cdot n_2$ that satisfy~(\ref{matta1})
with both values $x = 1$ and $x = 2$ is
$k_1 \cdot k_2$. Thus, the number of
states $q_w$, $w \in S_1$ for which the $b$-transition is defined is
$k_1 n_2 + k_2 n_1 - k_1 k_2$.

For $c^{m_1} b c^i \in S_2$, $0 \leq i < n_1$, the string
$c^{m_1} b c^i b$ is a prefix of some string of $L_1 \cup L_2$
if and only if $i \equiv j \; ({\rm mod} \; n_1)$,
$0 \leq j < k_1$. This means that the number of states
$q_w$, $w \in S_2$, for which the $b$-transition is
defined is $k_1$. Similarly, $S_3$ contains $k_2$ strings
$w$ such that the $b$-transition is defined for the
state $q_w$.

Putting the above together we have seen that
$$
\#_{\rm tr}(D, b) \geq k_1 n_2 + k_2 n_1 - k_1 k_2 + k_1 + k_2.
$$
Recalling that ${\rm sc}(L_i) = n_i$, ${\rm tc}_b(L_i) = k_i$,
$i = 1, 2$, the right side of the above inequality becomes
the right side of~(\ref{ubound}). Hence ${\rm tc}_b(L_1 \cup L_2)$
is exactly $k_1 n_2 + k_2 n_1 - k_1 k_2 + k_1 + k_2$.
\endpf

Note that the above lower bound construction does not work
with $k_i = n_i$, $i \in \{ 1, 2 \}$, because the proof relies on the property
that some $b$-transitions of $C_i$ are undefined.

For given relatively prime integers $n_1$ and $n_2$, the
construction used in Lemma~\ref{lem1} gives the maximum lower bound
for ${\rm tc}_b(L_1 \cup L_2)$ as a function of $n_i$ $(= {\rm
sc}(L_i))$, $i = 1, 2$, by choosing ${\rm tc}_b(L_i) = n_i-1$, $i =
1, 2$. In this case also ${\rm sc}(L_i) - {\rm tc}_b(L_i) = 1$.

On the other hand, by choosing $k_1 = k_2 = 1$, Lemma~\ref{lem1}
establishes that the $b$-transition complexity of $L_1 \cup L_2$ can
be arbitrarily larger than the $b$-transition complexity of the
languages $L_1$ and $L_2$. These observations are stated in the
below corollary.

\begin{corollary}
\label{cor4} Suppose that the alphabet $\Sigma$ has at least two
symbols and let $b \in \Sigma$ be a fixed symbol of $\Sigma$.
\begin{enumerate}
\item
For any relatively prime integers $n_1$ and $n_2$, there exist
regular languages $L_i$ with ${\rm sc}(L_i) =  n_i$, ${\rm
tc}_b(L_i) = n_i-1$,
 $i = 1, 2$, such that
$$
{\rm tc}_b(L_1 \cup L_2) = n_1n_2+n_1+n_2-3.
$$
\item For any constants  $h_i$, $i = 1, 2$, and $M \geq 1$
there exist regular languages $L_i$, $i = 1, 2$, such ${\rm
tc}_b(L_i) = h_i$, $i = 1, 2$, and ${\rm tc}_b(L_1 \cup L_2) \ge M$.
\end{enumerate}
\end{corollary}

For a given $b \in \Sigma$, Corollary~\ref{cor4}~(i) gives a lower
bound for ${\rm tc}_b(L_1 \cup L_2)$. The construction can be
extended for more than one alphabet symbol as indicated in
Corollary~\ref{cor5}, however, it cannot be extended to all the
alphabet symbols.

In the lower bound construction of the proof of Lemma~\ref{lem1},
the language $L_i$ was defined by a DFA that has a $c$-cycle of
length $n_i$, and where exactly $k_i$ of the states
had self-loops on symbol $b$. We can get
a simultaneous  lower bound for the number of $d$-transitions for
any $d \in \Sigma - \{ c, b \}$ by adding,
in a similar way, self-loops on the symbol $d$.

\begin{corollary}
\label{cor5}
Suppose that $\Sigma$ has at least two letters and fix $c \in
\Sigma$.
Let $n_1$ and $n_2$ be relatively prime
and for each $b \in \Sigma - \{ c \}$ fix a
number $1 \leq k_{i, b} < n_i$, $i = 1, 2$.

Then there exist regular languages $L_1$ and $L_2$ such that
$$
{\rm sc}(L_i) = n_i, \;\; {\rm tc}_b(L_i) = k_{i, b}, \;\;
b \in \Sigma - \{ c \}, \; i = 1, 2,
$$
and the equality holds in~(\ref{ubound}) for all $b \in \Sigma - \{ c \}$.
\end{corollary}

Finally we note that
the proof of Lemma~\ref{lem1} gives also the worst-case
bound for the state complexity of union for incomplete
DFAs.

\begin{corollary}
\label{cor3}
If ${\rm sc}(L_i) = n_i$, $i = 1, 2$, the language
$L_1 \cup L_2$ can be recognized by a DFA with at
most $n_1 \cdot n_2 + n_1 + n_2$ states. For relatively
prime  numbers $n_1, n_2 \geq 2$ the upper bound is tight.
\end{corollary}

\proof
The upper bound follows from the construction used in the
proof of Lemma~\ref{lem1}. The upper bound is reached
by the automata $A_1$ and $A_2$ used there for the
lower bound construction (with any values $1 \leq k_i < n_i$,
$i = 1, 2$).
\endpf

\subsection{Total number of transitions}

Here we give upper and lower bounds for
the  transition complexity
of
union of two regular languages.

With respect to the total number of  transitions for all input
symbols, the lower bound construction of the proof of
Lemma~\ref{lem1} maximizes ${\rm tc}(L_1 \cup L_2)$ as a function of
${\rm tc}(L_i)$, $i = 1, 2$, by choosing $k_1 = k_2 = 1$. In this
case it can be verified that ${\rm tc}(L_1 \cup L_2) = {\rm tc}(L_1)
\cdot {\rm tc}(L_2) + {\rm tc}(L_1) + {\rm tc}(L_2) - 2$. However,
when the alphabet has at least three symbols we can increase the
lower bound by one, roughly as in Corollary~\ref{cor5} by observing
that ${\rm tc}(L_i)$ can be chosen to be zero as long as for each $1
\leq i \leq 2$ there exists $b \in \Sigma$ such that ${\rm tc}(L_i)
\geq 1$. This is verified in the below lemma.

\begin{lemma}
\label{hatta1}
Let $\Sigma = \{ a, b, c \}$. For any relatively prime numbers
$n_1$ and $n_2$ there exist regular languages $L_i \subseteq \Sigma^*$,
 such that ${\rm tc}(L_i) = n_i + 1$, $i = 1, 2$, and
\begin{equation}
\label{ehat1}
{\rm tc}(L_1 \cup L_2) = {\rm tc}(L_1) \cdot {\rm tc}(L_2)
+ {\rm tc}(L_1) + {\rm tc}(L_2) - 1.
\end{equation}
\end{lemma}

Lemma~\ref{hatta1} can be proved with a construction similar to the
construction of the proof of Lemma~\ref{lem1} and witness languages
$L_1 = a^*(a^*c^{n_1})^*$ and $L_2 = b^*(b^*c^{n_2})^*$. Due to the
page limitation, we omit the proof. Next we give an upper bound for
transition complexity of union. In the following lemma let $A_1$ and
$A_2$ be arbitrary DFAs and $B_{A_1,A_2}$ denotes the DFA
constructed to recognize
 $L(A_1) \cup L(A_2)$  as in the proof of Lemma~\ref{lem1}.
(The definition of
$B_{A_1,A_2}$ is given in equation~(\ref{uconst}).)

\begin{lemma}
\label{satta1}
If  $A_i$ is connected,
$i = 1, 2$, then
$$
\#_{\rm tr}(B_{A_1,A_2}) \leq  2 \cdot (\#_{\rm tr}(A_1) \cdot \#_{\rm tr}(A_2)
+ \#_{\rm tr}(A_1) + \#_{\rm tr}(A_2)).
$$
\end{lemma}

\proof
We use induction on $\#_{\rm tr}(A_1) + \#_{\rm tr}(A_2)$.
First consider the case where $\#_{\rm tr}(A_1) =
\#_{\rm tr}(A_2) = 0$.
In this case also $B_{A_1,A_2}$ has no transitions.

Now assume that $\#_{\rm tr}(A_1) + \#_{\rm tr}(A_2) = m$,
and the claim holds when the total number of transitions
is at most $m-1$. Without loss of
generality, $\#_{\rm tr}(A_1) \geq 1$, and  let $A_1'$ be a connected DFA
obtained from $A_1$ by deleting one transition and possible states
that became disconnected as a result. We can choose the transition
to be deleted
in a way that at most one state becomes disconnected.

By the inductive hypothesis,
\begin{equation}
\label{esat3}
\#_{\rm tr}(B_{A_1', A_2}) \leq 2 \cdot
(\#_{\rm tr}(A_1') \cdot \#_{\rm tr}(A_2)
+ \#_{\rm tr}(A_1') + \#_{\rm tr}(A_2)).
\end{equation}
The DFA $A_1$ is obtained  by adding one transition $t_1$
and at most one state $q_1$ to $A_1'$.
Let $Q_2$ be the set of states of $A_2$.
The construction
of $B_{A_1,A_2}$ is the same as the
construction of $B_{A_1', A_2}$, except that
\begin{enumerate}
\item we add for $t_1$ a new transition
corresponding to each state of $Q_2$ and a new transition
corresponding to the dead state $d$ in the second component,  and,
\item we add a new transition corresponding to $q_1$ and each transition
of $A_2$.
\end{enumerate}
Thus,
$$
\#_{\rm tr}(B_{A_1, A_2}) \leq \#_{\rm tr}(B_{A_1', A_2})
+ |Q_2| + 1 +  \#_{\rm tr}(A_2) \leq  \#_{\rm tr}(B_{A_1', A_2}) +
2 (\#_{\rm tr}(A_2) + 1).
$$
The last inequality relies on~(\ref{esat1}) and the fact
that $A_2$ is connected.
Thus using~(\ref{esat3}) and $\#_{\rm tr}(A_1')
= \#_{\rm tr}(A_1) - 1$ we get
\begin{eqnarray*}
\#_{\rm tr}(B_{A_1, A_2}) & \leq  &
2((\#_{\rm tr}(A_1) - 1) \cdot \#_{\rm tr}(A_2) +
\#_{\rm tr}(A_1) - 1 + \#_{\rm tr}(A_2)) + 2 (\#_{\rm tr}(A_2) + 1).
\end{eqnarray*}
With arithmetic simplification this gives the claim for $A_1$
and $A_2$.
\endpf

\if01
\begin{lemma}
\label{hatta3}
Let $L_i$, $i = 1, 2$, be regular languages
with ${\rm tc}(L_i) \geq 1$. Then
\begin{equation}
\label{ehat3}
{\rm tc}(L_1 \cup L_2) \leq {\rm tc}(L_1) \cdot {\rm tc}(L_2) +
{\rm tc}(L_1) + {\rm tc}(L_2).
\end{equation}
\end{lemma}

\proof
We proceed by induction on ${\rm tc}(L_1) + {\rm tc}(L_2)$.
As the base case we consider
${\rm tc}(L_1) = {\rm tc}(L_2) = 1$.
All regular languages with transition complexity one
are $a^*$, $b$, and $\varepsilon + c$. It is easy to verify
that the union of any two of these languages (where symbols
$a$, $b$, $c$ may coincide) has a DFA with at most three transitions.

Inductively, assume that~(\ref{ehat3}) holds always
when ${\rm tc}(L_1) \cup {\rm tc}(L_2) \leq m (\geq 2)$.
Let $M_1$ and $M_2$ be regular languages such that
${\rm tc}(M_1) + {\rm tc}(M_2) = m+1$, (${\rm tc}(M_i) \geq 1$,
$i = 1, 2$), and let $A_i$ be the minimal DFA for $M_i$,
$i = 1, 2$.

Let $b_j$, $j = 1, \ldots, h$, be the elements of
$\Sigma$ that occur in words of $M_i$, $i = 1, 2$. (Note that
$A_i$ does not have any transitions on symbols that do
not occur in $M_i$. Denote
$$
k_i[j] = {\rm tc}_{b_j}(L_i), \;\; 1 \leq j \leq h, \; i = 1, 2.
$$
Note that for any $1 \leq j \leq h$, $k_1[j] \geq 1$ or $k_2[j] \geq 1$.

Without loss of generality suppose that ${\rm tc}(M_1) \geq 2$
and $M_1$ has a transition on symbol $b_h$. (This can always we
achieved by renaming the symbols $b_j$, $j = 1, \ldots, h$,
in a suitable way.) Let $B_1$ be a DFA that is obtained
from $A_1$ by deleting an arbitrarily chosen transition
on $b_h$ ($A_1$ has at least one transition on $b_h$).
Denote
$$
k_1'[j] = {\rm tc}_{b_j}(L(B_1))  \;\; 1 \leq j \leq h,
$$
that is, $k_1'[j]$ is the number of $b_j$ transitions in the
minimal DFA recognizing $L(B_1)$. In particular, we note that
${\rm tc}_{b_j}(L(B_1)) \leq \#_{\rm tr}(B_1)$ and it follows
that
\begin{equation}
\label{natta1}
k_1'[j] \leq k_1[j], \; 1 \leq j \leq h-1, \mbox{ and }
k_1'[h] \leq k_1[h] - 1.
\end{equation}
The above inequalities may not be equalities when $B_1$ is not
minimal.

Now we have
\begin{eqnarray}
\label{natta2}
{\rm tc}(L(B_1) \cup M_2)  =
\sum_{j = 1}^h {\rm tc}_{b_j}(L(B_1) \cup M_2) \nonumber \\
 \leq  \sum_{j=1}^h k_1'[j] k_2[j] +
 \sum_{j = 1}^h ( {\rm sc}(L_2) + 1 - k_2[j])k_1'[j] +
({\rm sc}(L(B_1)) + 1 - k_1'[j])k_2[j]) \nonumber \\
\leq \sum_{j=1}^h k_1'[j] k_2[j] + \nonumber
\end{eqnarray}
The first inequality uses Lemma~\ref{lem1} and the second
inequality uses the observation that for any regular language
$L$,
${\rm sc}(L) \leq $a 
\endpf
\fi

From Lemma~\ref{hatta1} and Lemma~\ref{satta1} we get now:

\begin{theorem}
\label{satta5}
For all regular languages $L_i$,
$i = 1, 2$,
$$
{\rm tc}(L_1 \cup L_2) \leq 2 \cdot ({\rm tc}(L_1) \cdot {\rm tc}(L_2)
+ {\rm tc}(L_1) + {\rm tc}(L_2)).
$$
For any relatively prime numbers $n_1$ and $n_2$ there exist regular
languages $L_i$ over a three-letter alphabet, ${\rm tc}(L_i) = n_i +
1$, $i = 1, 2$, such that
$$
{\rm tc}(L_1 \cup L_2)  = {\rm tc}(L_1) \cdot {\rm tc}(L_2)
+ {\rm tc}(L_1) + {\rm tc}(L_2) - 1.
$$
\end{theorem}

The upper and lower bound of Theorem~\ref{satta5} differ, roughly,
by a multiplicative constant of two. We believe that
the upper bound could be made lower
(when ${\rm tc}(L_i) \geq 2$, $i = 1, 2$), but do not have a proof
for this in the general case.

The constructions of Lemma~\ref{lem1} and Lemma~\ref{hatta1}
use languages $L_i$, $i = 1, 2$, such that for one particular
alphabet symbol $c \in \Sigma$, the minimal DFA for $L_i$, $i = 1, 2$,
has all $c$-transitions defined. It seems likely
that worst-case examples need to be based on cycles of transitions
on a particular alphabet symbol, in order to reach the
maximal state complexity blow-up with as small number of transitions
as possible. Below we establish that for this type
of constructions the right side of~(\ref{ehat1}) is also
an upper bound for ${\rm tc}(L_1 \cup L_2)$.


\begin{lemma}
\label{hatta2}
Let $L_1$ and $L_2$ be regular languages over $\Sigma$.
If there exists $c \in \Sigma$ such that in the minimal DFA
for $L_i$, $i = 1, 2$, all $c$-transitions are
defined, then
\begin{equation}
\label{ehat2}
{\rm tc}(L_1 \cup L_2) \leq {\rm tc}(L_1) \cdot {\rm tc}(L_2)
+ {\rm tc}(L_1) + {\rm tc}(L_2) - 1.
\end{equation}
\end{lemma}

The idea of the proof of Lemma~\ref{hatta2} is similar to that of
the proof of Lemma~\ref{satta1}. The crucial difference is that we
have one symbol for which all transitions are defined and the
inductive argument is with respect to the number of the remaining
transitions. Thus, in the inductive step when replacing $A_1$ with a
DFA $A_1'$ with one fewer transition,  we know that $A_1'$ is
connected and the inductive step does not need to add transitions
corresponding to a state that would be added to $A_1'$.

Lemma~\ref{hatta2} establishes that the bound given by
Lemma~\ref{hatta1} cannot be exceeded by any construction that is
based on automata that both have a complete cycle defined on the
same alphabet symbol. Usually it is easier to establish upper bounds
for descriptional complexity measures, and finding matching lower
bounds is a relatively harder question. In the case of transition
complexity of union we have a lower bound and only indirect
evidence, via Lemma~\ref{hatta2}, that this lower bound cannot be
exceeded.

\begin{conjecture}
For any regular languages $L_1$ and $L_2$ where
${\rm tc}(L_i) \geq 2$, $i = 1, 2$,
$$
{\rm tc}(L_1 \cup L_2) \leq {\rm tc}(L_1) \cdot {\rm tc}(L_2) + {\rm
tc}(L_1) + {\rm tc}(L_2).
$$
\end{conjecture}

Note that the conjecture does not hold for small values of ${\rm
tc}(L_i)$, $i = 1, 2$. For example, ${\rm tc}(\{ \varepsilon \}) =
0$, ${\rm tc}(a^* b^{m-1}) = m$, but ${\rm tc}(a^* b^{m-1} +
\varepsilon) = m+2$.

\subsection{Transition complexity of union of unary languages}

For languages over a unary alphabet, the transition complexity
of union of incomplete DFAs turns out to coincide with  the
known bound for state complexity of union of complete DFAs.
However, the proof  is slightly different.

Recall that a DFA with a unary input alphabet
always has a ``tail'' possibly
followed by a ``loop''~\cite{Ch,PiSh2002}. Note that an incomplete
DFA recognizing a finite language does not need to have a loop.

\begin{theorem}
\label{lem2}
Let $L_1, L_2$ be unary
languages over an alphabet $\{ b \}$. If ${\rm tc}(L_i) \geq 2$,
$i = 1, 2$, then
\begin{equation}
\label{ubound2}
{\rm tc}(L_1 \cup L_2) \leq {\rm tc}(L_1) \cdot {\rm tc}(L_2).
\end{equation}
For any relatively prime $n_1\geq 3$, $n_2 \geq 2$, there exist
regular languages $L_i \subseteq \{ b \}^*$, ${\rm tc}(L_i) = n_i$,
$i = 1, 2$, such that~(\ref{ubound2}) is an equality.
\end{theorem}




This theorem can be proved by separately considering the cases where
$L_1$ and $L_2$ are finite or infinite. The detailed proof is
omitted. Note that the upper bound of Theorem~\ref{lem2} does not
hold when $n_1<3$ or $n_2 < 2$. For example, ${\rm tc}(b) = 1$,
${\rm tc}((b^n)^*) = n$ and ${\rm tc}( b \cup (b^n)^*) = n+1$ when
$n \geq 2$.

\section{Intersection and complementation}

As can, perhaps,  be expected the worst-case transition complexity
bounds for intersection are the same as the corresponding state complexity
results based on complete DFAs. When dealing with intersection,
worst-case examples can be constructed using a unary alphabet
and a complete DFA. On the other hand, state complexity of complementation
of complete DFAs is the identity function whereas the bound
for transition complexity of complementation
is significantly different.

\begin{proposition}
\label{iprop}
For any regular languages $L_i$, $i = 1, 2$,
\begin{equation}
\label{fatta1}
{\rm tc}(L_1 \cap L_2) \leq {\rm tc}(L_1) \cdot {\rm tc}(L_2).
\end{equation}
Always when $n_1$ and $n_2$ are relatively prime there exist
regular languages $L_i$, ${\rm tc}(L_i) = n_i$, $i = 1, 2$, such
that equality holds in~(\ref{fatta1}).
\end{proposition}

\proof
Let  $A_i = (\Sigma,
Q_i, q_{0,i}, F_i, \delta_i)$ be a DFA
 recognizing
$L_i$, $i = 1, 2$.
We define
\begin{equation}
\label{iconst}
B = (\Sigma, Q_1 \times Q_2, (q_{0,1}, q_{0,2}), F_1 \times F_2,
\gamma),
\end{equation}
where for $b \in \Sigma$, $q_i \in Q_i$, $i = 1, 2$,
$$
\gamma((q_1, q_2), b) = \left\{
\begin{array}{l}
(\delta_1(q_1, b), \delta_2(q_2, b)) \mbox{ if }
\delta_1(q_1, b) \mbox{ and }
\delta_2(q_2, b) \mbox{ are both defined, } \\
\mbox{undefined, otherwise.}
\end{array} \right.
$$
Clearly $B$ recognizes $L_1 \cap L_2$ and
$\#_{\rm tr}(B) = \#_{\rm tr}(A_1) \cdot \#_{\rm tr}(A_2)$.

The lower bound follows from the observation that if
 $n_1$ and $n_2$ are relatively prime and $A_i$ is the
minimal DFA for $(b^{n_i})^*$, $i = 1, 2$, then
the DFA $B$ in~(\ref{iconst}) is also minimal and
$\#_{\rm tr}(B) = n_1 n_2$.
\endpf

The proof of Proposition~\ref{iprop} gives
for $b \in \Sigma$  the same tight bound
for the number of $b$-transitions needed to recognize the
intersection of given languages.

\begin{corollary}
\label{cor20}
For any regular languages $L_i$ over $\Sigma$,
$i = 1, 2$, and $b \in \Sigma$,
\begin{equation}
\label{matta3}
{\rm tc}_b(L_1 \cap L_2) \leq {\rm tc}_b(L_1) \cdot {\rm tc}_b(L_2).
\end{equation}
For  relatively prime integers $n_1$ and $n_2$ there exist regular
languages $L_i$ with ${\rm tc}_b(L_i) = n_i$, $i = 1, 2$,
such that equality holds in~(\ref{matta3}).
\end{corollary}

To conclude this section we consider complementation.
If $A$ is an $n$-state DFA, a DFA to recognize the complement of
$L(A)$ needs at most $n + 1$ states.
The worst-case bound for transition complexity
of complementation is significantly
different.

\begin{proposition}
\label{cprop}
Let $L$ be a regular language over an alphabet $\Sigma$.
The transition complexity of
the complement of $L$ is upper bounded by
$$
{\rm tc}(L^c) \leq |\Sigma| \cdot ({\rm tc}(L) + 2).
$$
The bound is tight, that is, for any $n \geq 1$ there exists
a regular language $L$ with ${\rm tc}(L) = n$ such that
in the above inequality
the equality holds.
\end{proposition}

\proof
Let $A = (\Sigma, Q, q_0, F, \delta)$ be a DFA for $L$.
The complement of $L$ is recognized by the
DFA
$$B = (\Sigma, Q \cup \{ d \}, q_0, (Q - F) \cup \{ d \},
\gamma),
$$
where for $b \in \Sigma$
$$
\gamma(p, b) = \left\{
\begin{array}{l}
\delta(p, b) \mbox{ if } p \in Q \mbox{ and }
\delta(p, b) \mbox{ is defined,}\\
d \mbox{ if } \delta(p, b) \mbox{ is undefined.}
\end{array} \right.
$$
Note that when $p = d$, $\delta(p, b)$ is
undefined for all $b \in \Sigma$.

The DFA $B$ has $(|Q| + 1) \cdot |\Sigma|$ transitions.
If $A$ is minimal, $A$ has at least $|Q| - 1$ transitions,
and this gives the upper bound.

We establish the lower bound. Choose $b \in \Sigma$ and for
$n \geq 1$ define $L_n = \{ b^n \}$. Now
${\rm tc}(L_n) = n$. Denote $S = \{
\varepsilon, b, \ldots, b^{n+1} \}$. All strings
of $S$ are
pairwise inequivalent with respect to the right
congruence $\equiv_{L_n^c}$, and
$$
(\forall x \in S)(\forall c \in \Sigma)(\exists y \in \Sigma^*)
\;\; xcy \in L_n^c.
$$
 This means that the minimal DFA for $L_n^c$ has
(at least) $n+2$ states for which all transitions are
defined. Thus, ${\rm tc}(L_n^c) \geq |\Sigma| \cdot (n+2)$.
\endpf

From the construction of the proof of Proposition~\ref{cprop} we see
that if $\Sigma$ contains at least two symbols then for $a \in
\Sigma$ and any $M \geq 1$ there exists a regular language $L$ over
$\Sigma$ such that ${\rm tc}_a(L) = 0$ and ${\rm tc}_a(L^c) \geq M$.


\section*{Acknowledgement} We would like to thank the anonymous
referees of $DCFS2010$ for their careful reading and valuable
suggestions.

\end{document}